# THERMOELECTRIC POWER OF Tl-DOPED PbTe MONOCRYSTAL

E. A. Zasavitsky

International Laboratory of Solid State Electronics (LISES),
Moldavian Academy of Sciences.
Academiei str.3/3, Chisinau, MD-2028, Moldova.
E-mail: efim@lises.asm.md

**Abstract:** Results of the measurements of thermoelectric properties of thin semiconductor microwires of $Pb_{1-x}Tl_xTe$ (x=0.001 ÷ 0.02, d = 5 ÷ 100 μm) in the temperature region 4,2 ÷ 300 K, which were obtained from solution melt by the filling of quartz capillary with the following crystallization of material are presented. For the samples corresponding to chemical composition with concentration of thallium $0,0025 < x < 0,005$ double change of the sign of thermoelectric power is observed. In pure samples and samples with thallium concentration more than 1 at.% thermoelectric power it is positive in the whole temperature range. Various mechanisms which can lead to observable anomalies, including Kondo-like behavior of a non-magnetic degenerate two-level system are discussed. Obtained experimental results let suppose that the observed anomalies can be interpreted on the basis of model of an impurity with mixed valences.

## 1. INTRODUCTION.

Semiconductor compounds of type $A_4B_6$ doped by elements of III group manifest a number of unusual features in their electro physical and optical properties [1, 2] useful for different applications, in particular for thermoelectric cooling. By that is explained the interest to such materials and, in particular, to semiconductor compound $Pb_{1-x}Tl_xTe$ as to the most striking example in which a lot of unique properties have been observed. Specific action of thallium consists in the following: the thallium impurity generates on the range of the permitted band states of a valence band an impurity band that leads to essential change of density of band states. It leads to essential change of transport characteristics which are determined by density of states, by the form of the impurity band and by the relative position of the Fermi level [1, 2]. On the other hand, thallium is an impurity elements of III group in the compounds connections $A_4B_6$ which show a mixed valence [3]. It means that thallium can have the valence from 1 up to 3, and the most unstable is the state with a bivalent impurity. In such case they should dissociate as [4]:

$$2Tl^{2+} \rightarrow Tl^{1+} + Tl^{3+}$$

Theoretical description of that assumes existence of the centers with negative correlation energy U:

$$U_n = (E_{n+1} - E_n) - (E_n - E_{n-1}) < 0,$$

where valence number n takes the values 1, 2, 3.

Theoretically it has been predicted, that the presence of the U-centers can lead to those features in electro physical properties which were observed in the system $Pb_{1-x}Tl_xTe$ [4]. Such a behavior of an impurity of thallium in PbTe is similar to Kondo effect in metals and should be manifests in particular on the thermoelectric properties of this compound. In the presented work the results of experimental study of thermoelectric properties of $Pb_{1-x}Tl_xTe$ semiconductor microwires are presented and analyzed on the basis of the model of resonant impurity states with negative correlation energy.

## 2. EXPERIMENTAL RESULTS AND DISCUSSION.

Semiconductor microwires of $Pb_{1-x}Tl_xTe$ (diameter d = 5 ÷ 100 μm, length l ~ 20 sm) with thallium average concentration x = 0.001 ÷ 0.02 were grown in the following way [5] (Fig. 1). In the quartz tube (diameter - 15 mm) initial material with corresponding chemical composition was placed. The bulk material was prepared in the following way. Since pure Tl oxidizes in the air quickly and greatly, it is necessary for a preparation of initial mixtures to use compounds of thallium - in our case TlTe. Syntheses of polycrystalline materials $(PbTe)_{1-x}(TlTe)_x$ of corresponding compounds were made in the quartz tube in the hydrogen atmosphere. Over the material the bunch of quartz capillary is situated. The choice of quartz as the material for capillaries is limited by the high temperature of softening one, what must be higher than the melting temperature of material. The tube was evacuated up to residual pressure $10^{-2} \div 10^{-3}$ Pa and placed in vertical zone furnace, in which the temperature on the whole length of the capillary is the same and higher than the melt temperature of material ($T_{melt} < T < T_{soft}$). After melting of material the capillary with open lower ends were put down in the melt material. Afterwards in the tube rise pressure under which capillary were filled by the melting material. Crystallization of melting material was realized directly beginning from soldered ends to open one at the expense of move of furnace (rate of move may be changed and make up several centimeters per hour). Given method of obtaining of monocrystal microwires allows to produce samples with different diameters under the same grown conditions with high structural perfection. The structural quality was tested by X-ray diffraction and Laser Microprobe Mass Analyzer (LAMMA) (Fig. 2).

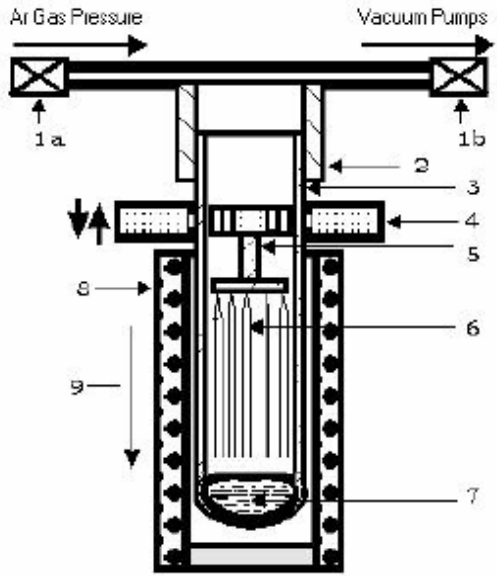

Fig.1. Sketch of the laboratory-scale apparatus for fabrication of thin glass-coated semiconducting wires using the high-pressure injection and directional crystallization method.
1a, 1b - vacuum valves; 2 - metallic tube; 3 - quartz tube; 4 - permanent-magnet system to move capillaries; 5- support for capillaries; 6 - glass capillaries; 7 - molten material; 8 - electric furnace; 9 - direction of furnace movement during wire crystallization.

The samples for the measurements were prepared in the following mode. The sample of the corresponding diameter was choused from the set of crystals obtained in that way for carrying out of measurements. As the initial sample has glass isolation, it was preliminary subjected to selective etching in a solution of acid HF. Reliable electrical contact was made using eutectic In-Ga. Measurements have been executed both on samples with different diameters, and on samples of the same diameters. Measurements of temperature were carried out by means of thermocouple Cu – (Cu + 0,04 at % Fe). Such thermocouple make possible to carry out experiments with high precision in low temperature region.

In the Fig.3, 4 temperature dependences of thermoelectric power of monocrystals microwires of $Pb_{1-x}Tl_xTe$ are shown. The analysis of temperature dependences of thermoelectric power shows, that in the doped samples thermoelectric power manifested anomalous character - at low temperatures this dependence becomes essentially non monotonic. For $Pb_{1-x}Tl_xTe$ (x=0,0025; 0,005) thermoelectric power changes the sign twice, and for other compounds on temperature dependence the curve without change of the sign is shown. At higher temperatures (T > 100K) dependence of thermoelectric power vs temperature shows usual behavior, characteristic for strongly doped lead telluride, and numerical value of thermoelectric power in this region are comparable with the data resulted in [6].

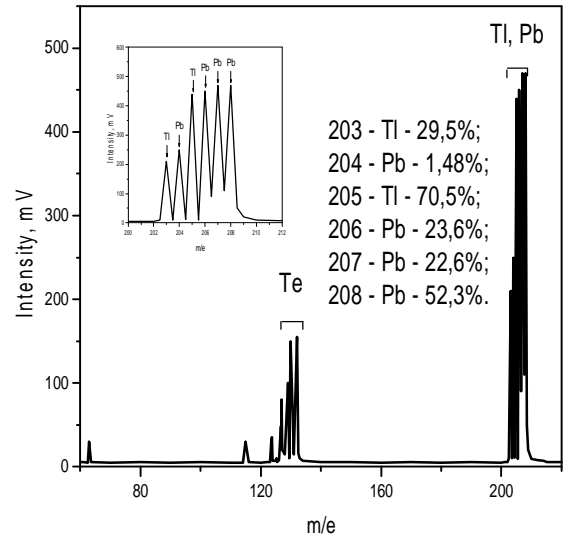

Fig.2. Characteristic LAMMA spectra of monocrystal wires of $Pb_{1-x}Tl_xTe$.

It is known, that in strong degenerated samples the value of the coefficient of thermoelectric power are determined by parameter of dispersion $r$.

The change of the sign of thermoelectric power in the dependence of concentration of thallium impurity at temperatures above the temperature of liquid nitrogen in lead telluride was observed earlier [7] and was interpreted within the model of features of resonant scattering of carriers. On dependence of thermoelectric power from concentration of thallium the deep minimum downing up to change of its sign [2] is observed. For such behavior of thermoelectric power the expression for the parameter of scattering in the conditions of resonant scattering describing domination was proposed in the form:

$$r = \frac{2m(m - e_i)}{(m - e_i)^2 + (\Gamma/2)^2},$$

where $m$ is chemical potential, $e_i$ is the mid position the impurity band and $\Gamma$ its width.

Theoretical calculations are done and they agreed with experimental results for a lot of $A_4B_6$ compounds doped with Tl. However as follows from obtained experimental data the change of the sign of thermoelectric power is observed both at high and low temperatures. Furthermore the resonant scattering of carriers is not a process of activation type. It means, that at $k_0T \ll \Gamma$ relaxation time does not depend on temperature. Therefore the anomaly behavior of thermoelectric power at low temperatures cannot be explained with features of resonant scattering.

The reason of low temperature changes of the sign of thermoelectric power can be electron-phonon drag. It was observed earlier for pure lead telluride at T <20K (p = $10^{18}$

$$a = \frac{a_1\sqrt{s_1 k_1} + a_2\sqrt{s_2 k_2}}{\sqrt{s_1 k_2} + \sqrt{s_2 k_1}},$$

where $\kappa_1$, $\kappa_2$ are the thermal conductivities of the phases; $\sigma_1$, $\sigma_1$ are the conductivities of the phases, $\alpha_1$, $\alpha_2$ are their thermoelectric powers.

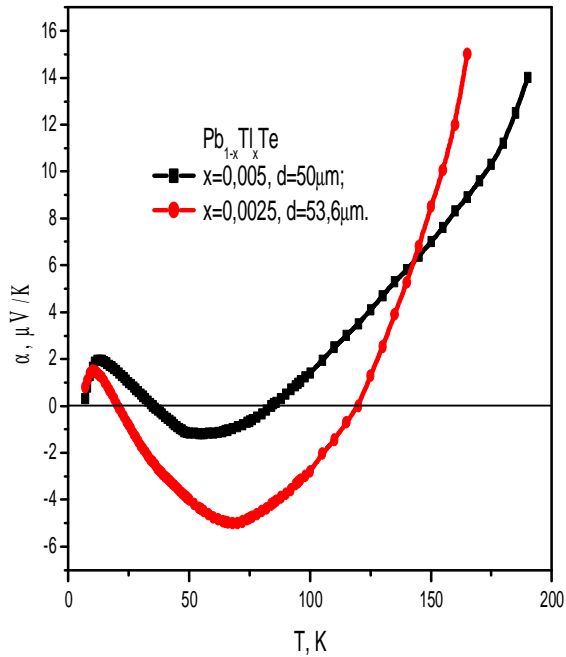

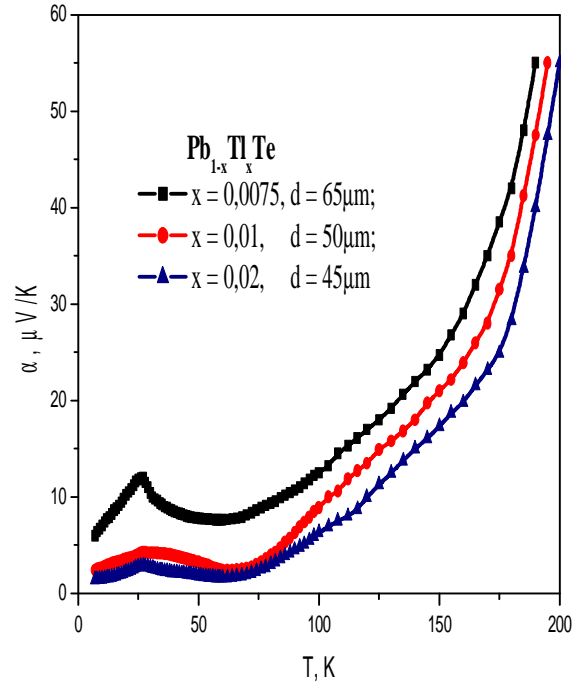

Fig.3. Temperature dependences of thermoelectric power of monocrystal wires of $Pb_{1-x}Tl_xTe$.

Fig.4. Temperature dependences of thermoelectric power of monocrystal wires of $Pb_{1-x}Tl_xTe$.

sm$^{-3}$) [8]. However in monocrystal microwires of $Pb_{1-x}Tl_xTe$ where concentration of carriers is more than on the order larger than in samples investigated in [8] electron-phonon drag cannot lead to low temperature change of thermoelectric power sign.

The assumption of the mixed mechanism of scattering of carriers also does not allow explaining the no monotonic temperature dependence of thermoelectric power.

Other mechanism, enabling an adequate explanation of observed anomalies of temperature dependences of thermoelectric power, is based on the theory of the transport phenomena in large scale non-uniform medium. Large scale non-uniform is possible to count a medium, in which the size of heterogeneity is much greater then the free path lengths in each of phases. In our case it is necessary to understand areas with various values of local kinetic factors.

Qualitative results for effective thermoelectric power for quasi-two-dimensional two-phase thin monocrystal microwires are received [9]:

At thallium concentration in lead telluride corresponding to the beginning of overlapping of Fermi level with Tl-impurity band, small fluctuations in distribution of the impurities should lead to the occurrence of the areas with sharply distinguished transport parameters due to threshold character of resonant scattering. Thus having attributed the index 1 for the areas with low impurity concentration (Fermi level outside of Tl-impurity band), and the index 2 for the areas with Fermi's level in Tl-impurity band we can describe qualitatively abovementioned kinds of temperature dependence of thermoelectric power. Because at low temperatures thermal conductivity has mainly lattice character small fluctuations of the impurity concentration will not considerably affect their values, that is $\kappa_1 = \kappa_2$ and

$$a = a_1\left[1 + \frac{4}{3}\frac{m\Delta}{\Delta^2 + (\Gamma/2)^2}\left(1 + \sqrt{\frac{s_1(0)}{s_2(1+AT)}}\right)\right],$$

where A is an fitting parameter determined from temperature dependences of σ(T) for nonstoichiometry thin monocrystal wires of PbTe. It is established, that settlement curves qualitatively reproduce observed kind of temperature dependences of thermoelectric power in the range of low temperatures that testifies for the argumentation of the assumptions made above.

It is necessary to note, that the assumption about large-scale non-uniform medium is based on the experimental results obtained earlier for monocrystals wires of $Pb_{1-x}Tl_xTe$ [10].

In spite of this it is necessary to mention, that the set of results obtained on massive crystals indicate that observed anomalies in electrophysical and thermoelectrical properties cannot be explained only by non-uniform distribution of an impurity. The identified anomalies have their intrinsic distinctive features of the given material - lead telluride doped by thallium. Non-uniform distribution of an impurity which is most pronounced shown in the low dimensional systems, can increase (or reduce) only the manifestation of the effects which origin can to be connected with features of impurity Tl in PbTe which influence is similar to influence of magnetic impurity in metals.

The analysis of temperature dependences of thermoelectric power of semiconductor microwires of $Pb_{1-x}Tl_xTe$ in all range of concentration of a doping impurity at low temperatures shows, that the appearance of the maximum is characteristic. It is known [11], that in systems with magnetic impurity the appearance of low temperature maximum of thermoelectric power is provided by the interaction of of the carriers with the magnetic moments of impurities. Isolated Kondo-impurity gives a negative contribution in thermoelectric power and as a first approximation does not depended neither on temperature, nor from concentration of the Kondo-centers. From concentration of the impurities depends only the temperature at which thermoelectric power starts to deviate from the behavior characteristic for metals. Such picture is observed until then while the temperature will not go down to Kondo temperature where thermoelectric power starts to decrease quickly on the module, tending to zero value at T → 0K.

It is necessary to add, that in investigated monocrystal microwires of $Pb_{1-x}Tl_xTe$ the whole of characteristic electrophysical properties for bulk crystals [12] are observed:

- superconductivity with $T_c > 0.4$ K x > 0.01;
- starting with T = 30÷40 K samples exhibited negative temperature coefficients of resistance down to T = 0.4 K;
- all samples demonstrated negative magnetoresistance at helium temperature.

## 3. Conclusions.

The work includes the results on research of thermoelectric properties of semiconductor microwires of $Pb_{1-x}Tl_xTe$ (x=0.001 ÷ 0.02, d = 5 ÷ 100 μm) in a wide range of temperatures are presented. The main task was to identify new ways to modify the thermoelectric properties of such systems through resonant impurity states with negative correlation energy. For all investigated at low temperatures anomalies on the temperature dependence of thermoelectric power are observed. Taking into account features of thallium impurity in lead telluride and comparing the obtained results with similar for metals with magnetic impurity we conclude, that the nature of observed anomalies is the charge Kondo effect associated with mixed valence of Tl impurity states, which manifested in the PbTe.

Similar conclusions about features of doping action of thallium in monocrystals of $Pb_{1-x}Tl_xTe$ (Tl concentrations up to 1,5 at. %) was made in work [13] on the basis of the executed measurements of thermodynamic and transport properties in low-temperature region. Moreover, Kondo-type behavior of impurity of thallium in PbTe is that mechanism which allows explaining the anomalous high temperature of superconducting transition in Tl-doped PbTe.

It is necessary to note, that the question on a charging state of thallium in lead telluride remaining opened in comparison with other impurities of elements of III group [3]. Set of available experimental data on thallium impurity in PbTe are explained on the basis of model of occurrence on the background of the permitted band states of a valence band of a resonant impurity level with high density of states near the valence band [1,2]. It is possible to expect, that in connection with growth of interests to compounds $A_4B_6$ doped with elements of III group in the near future will be possible by direct experimental methods to establish character of behavior of Tl impurity in PbTe. The observed behavior of thermoelectric properties under conditions of resonant impurity states with mixed states offer new possibilities for reengineering of thermoelectric transport characteristics of semiconductors.

## 4. References.

[1] Kaidanov V. I., Ravich Y. I., Sov. Phys. Usp, **145,** 51, 1985.
[2] Nemov S. A., Ravich Y. I., Sov. Phys. Usp, **168,** 817, 1998.
[3] Volkov B. A., Ryabova L. I., Khokhlov D. R., Sov. Phys. Usp, **172, 875,** 2002.
[4] Drabkin I. A., Moizhes B. Y., Sov. Phys. Semicond. **15,** 357, 1981.
[5] Leporda N. I., Grozav A. D., Moldavian Journal of the Physical Sciences, **1,** 74, 2002.
[6] Cernik I. A., Kaidanov V. I., Vinogradova M. I., Kolomoets I. L., Sov. Phys. Semicond. **2,** 773, 1968.
[7] Veis A. N., Nemov S. A., Sov. Phys. Semicond, **15,** 1237, 1981.
[8] Regel L. L., Rahmatov O. I., Redko N. A., Sov. Phys. Solid State. **26, 1242,** 1984.

[9] Boiko M. P., Zasavitsky E. A. Sov. Phys. Semicond, **26,** 168, 1992.
[10] Boiko M. P., Ghitsu D. V., Zasavitsky E. A., Sidorenko A. S., Sov. Phys. Semicond, **21,** 1303, 1987.
[11] Abrikosov A. A. Bases of the theory of metals, Nauka, Moskow, p.91.
[12] Andronik K. I., Banar V. F., Kantser V. G., Sidorenko A. S., Phys. Stat. Sol. (b), **133,** K61, 1986.
[13] Matsushita Y., Bluhm H., Geballe T. H., Fisher I. R., Phys. Rev. Lett., **94,** 157002 (2005).